\documentclass[twocolumn]{aastex6}
\usepackage{color,amsmath,textcomp,url,graphicx,subfigure,latexsym}

\newcommand{\kms}{\hbox{km\,s$^{-1}$}}

\newcommand{\MJup}{$M_{\mathrm{Jup}}$}

\newcommand{\masyr}{$\mathrm{mas}\,\mathrm{yr}^{-1}$}
\newcommand{\Teff}{\ensuremath{T_{\mathrm{eff}}}}

\newcommand{\targetfull}{2MASS~J11193254--1137466}
\newcommand{\target}{2MASS~J1119--1137}

\shorttitle{Nearest TWA Member is a Giant Planet Analog}
\shortauthors{Kellogg et al.}

\begin{document}

\title{The Nearest Isolated Member of the TW Hydrae Association is a Giant Planet Analog}
\author{Kendra Kellogg\altaffilmark{1}, Stanimir Metchev\altaffilmark{1,2}, Jonathan Gagn\'{e}\altaffilmark{3,5}, \and Jacqueline Faherty\altaffilmark{3,4,6}}
\altaffiltext{1}{The University of Western Ontario, Centre for Planetary and Space Exploration, Department of Physics \& Astronomy, 1151 Richmond St, London, ON N6A 3K7, Canada; kkellogg@uwo.ca}
\altaffiltext{2}{Stony Brook University, Department of Physics \& Astronomy, 100 Nicolls Rd, Stony Brook, NY 11794-3800, USA}
\altaffiltext{3}{Carnegie Institution of Washington DTM, 5241 Broad Branch Road NW, Washington, DC~20015, USA}
\altaffiltext{4}{Department of Astrophysics, American Museum of Natural History, Central Park West at 79th Street, New York, NY 10034, USA}
\altaffiltext{5}{NASA Sagan Fellow}
\altaffiltext{6}{Hubble Fellow}

\begin{abstract}

In a recent search for unusually red L and T dwarfs, we identified \targetfull\ as a likely young L7 dwarf and potential member of the TW~Hydrae association. We present spectra that confirm the youth of this object. We also measure a radial velocity of $8.5 \pm 3.3$\,\kms\ that, together with the sky position, proper motion and photometric distance, results in a 92\% probability of membership in the TW~Hydrae association, with a calibrated field contamination probability of 0.0005\% using the BANYAN II tool. Using the age of TW~Hydrae and the luminosity of \targetfull, we estimate its mass to be 4.3--7.6\,\MJup.  It is the lowest-mass and nearest isolated member of TW~Hydrae at a kinematic distance of $28.9 \pm 3.6$\,pc, and the second-brightest isolated $<$10\,\MJup\ object discovered to date.

\end{abstract}

\keywords{brown dwarfs --- open clusters and associations: individual (TW~Hydrae) --- techniques: radial velocities --- stars: individual (2MASS~J11193254$-$1137466)}

\section{INTRODUCTION}

Young brown dwarfs, especially at the latest spectral types, have masses and atmospheres similar to those of directly imaged gas giant exoplanets. Isolated young brown dwarfs offer a way to study cool, low-pressure atmospheres of exoplanets without the inherent difficulties of isolating the planet flux from that of a brighter host star. 
Most of the known isolated planetary-mass brown dwarfs have been found through their unusually red optical and near-infrared colors, often in the regions of young stellar associations \citep[e.g.,][]{luhman_etal05c, marsh_etal10, aller_etal13, 2013ApJ...777L..20L, schneider14}.  Over the past few years, targeted searches have also encompassed the position-velocity phase spaces of nearby young stellar associations \citep[e.g.,][]{2014ApJ...783..121G, gagne15a, gagne15c}, most notably with the Bayesian Analysis for Nearby Young AssociatioNs (BANYAN) tool \citep{malo_etal13}.  These have helped recognize or discover the lowest-mass isolated brown dwarfs in the solar neighborhood \citep{delorme_etal12, 2013ApJ...777L..20L, 2014ApJ...785L..14G, 2015ApJ...808L..20G}.  

In \cite{2015AJ....150..182K} we conducted a comprehensive program to purposefully seek L and T dwarfs with unusual optical/infrared colors in the combined set of SDSS, 2MASS, and WISE data.  We did not impose positional or space velocity constraints, in order to obtain an unbiased assessment of ultra-cool dwarfs with peculiar spectral energy distributions (SEDs).  One of the newly discovered objects was the extremely red ($J-K_s=2.58\pm0.03$~mag) L7 candidate low-gravity dwarf 2MASS~J11193254$-$1137466 (henceforth, \target).  At $K_s=14.75\pm0.01$~mag in the VISTA Hemisphere Survey (VHS; PI McMahon, Cambridge, UK), a confirmation of its youth would place it among the brightest isolated planetary-mass objects, making it exceptionally suitable for investigating exoplanetary atmospheres in detail.

Herein we report the spectroscopic confirmation of the low surface gravity of \target, its kinematic association with the TW Hydrae association (TWA; \citealp{webb_etal99}), and hence its planetary mass.

\section{OBSERVATIONS and DATA REDUCTION}\label{sec:obs}

To confirm the youth of \target, we obtained low-resolution spectra with FLAMINGOS-2 \citep{f2} on the Gemini-South telescope through Fast Turnaround program GS-2015B-FT-5. The spectrum revealed weaker gravity-sensitive K I absorption lines than in a typical field-age late-L dwarf indicating that \target\ is young (e.g. \citealp{cruz09,rice10}). Because of unresolved spectrophotometric systematics, namely an extremely red spectral slope and unexpectedly deep OH absorption bands inconsistent with our SpeX spectrum \citep{2015AJ....150..182K}, we do not present the F-2 spectrum here. We subsequently obtained an $R$\,$\sim$\,6000, 0.8--2.45\,$\mu$m spectrum with the Folded-port InfraRed Echellette (FIRE; \citealp{2008SPIE.7014E..0US,2013PASP..125..270S}) at the Magellan Baade telescope on 2016 January 21. It is this spectrum that we use for the surface gravity and radial velocity analysis presented herein. 
We observed the target at the parallactic angle (PA) $\sim$\,226\textdegree, at an airmass of 1.17--1.05 with good weather conditions and a seeing of $\sim$\,0\farcs5. The echelle disperser was used in conjunction with the Sample Up the Ramp (SUTR) readout and high-gain (1.3\,e-/DN) modes. A single 900\,s exposure and six 600\,s exposures were obtained for a total integration time of 1.25\,hr, which yielded a signal-to-noise ratios of $\sim$\,10--70 per pixel in the 1.1--2.2\,$\mu$m range. We observed the A0 standard HD~85056 immediately afterwards at a similar airmass (1.08--1.09).

We used ThAr calibration lamps (10\,s exposure times), observed in the middle and at the end of the science sequence, and again after the  telluric standard, to ensure a proper wavelength calibration. A total of eleven internal high-signal flats (1\,s exposures with high-voltage lamps) and eleven low-signal flats (10\,s exposures with low-voltage lamps) were obtained at the beginning of the night: high-signal flats provide a good pixel response calibration in the blue orders, whereas low-signal flats are used for the red orders. Five 4\,s dome flats were also obtained to characterize the slit illumination function.

We reduced the data using the Interactive Data Language (IDL) pipeline FIREHOSE, which is based on the MASE \citep{2009PASP..121.1409B} and SpeXTool \citep{2003PASP..115..389V, 2004PASP..116..362C} packages. The pipeline was modified to achieve a better rejection of bad pixels (i.e., see \citealp{gagne15c,zenodofirehose}\footnote{The modified FireHose~2.0 data reduction package can be found at \url{https://github.com/jgagneastro/FireHose\_v2/tree/v2.0}}). This reduction package includes a barycentric velocity correction and converts the wavelength solution to vacuum.

\section{CONFIRMATION OF YOUTH}\label{sec:youth}

\begin{figure*} 
	\centering 
	\includegraphics[scale=0.35]{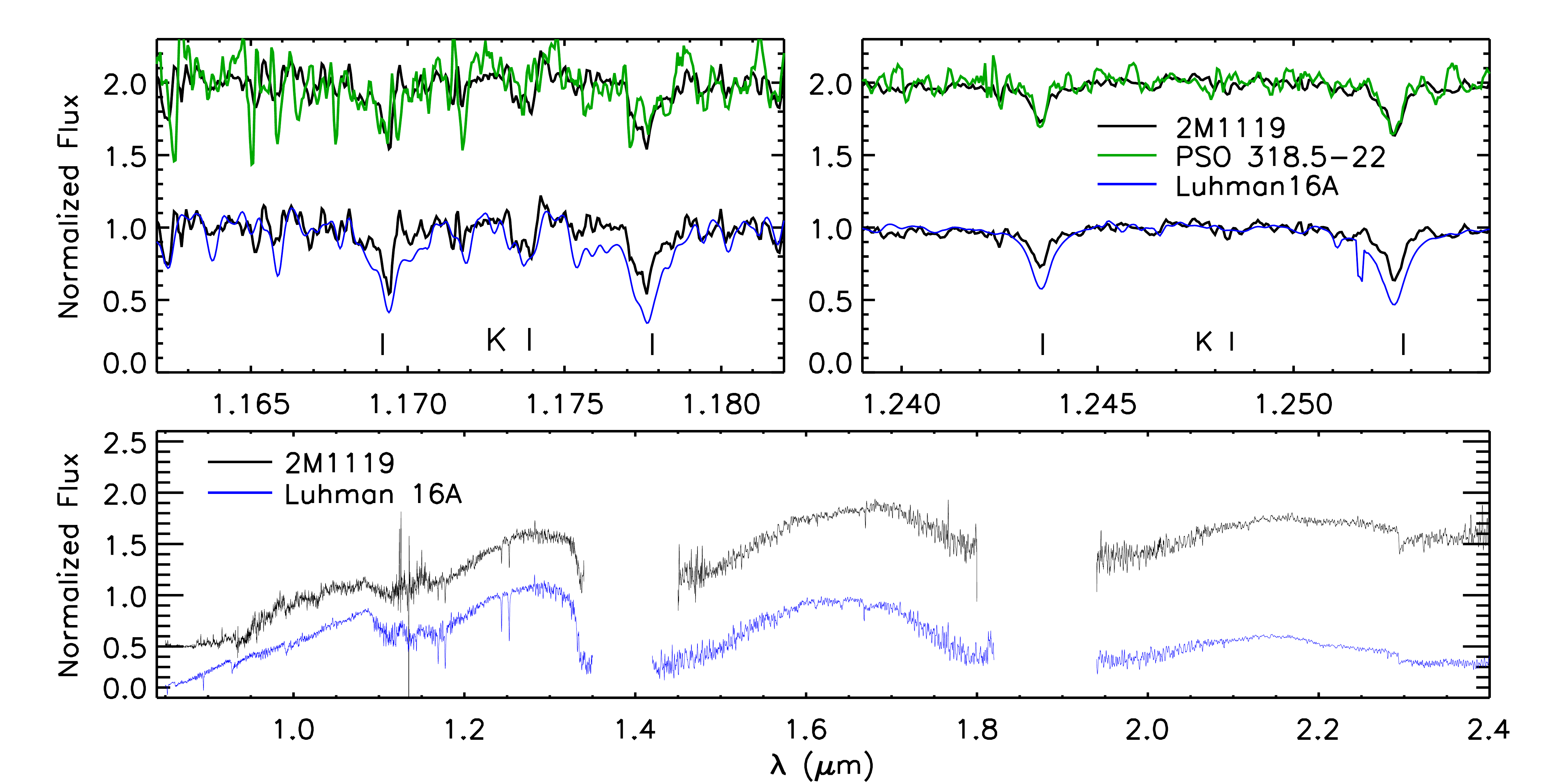}
	\caption{Comparison of the $R\sim6000$ FIRE spectra of \target\ (black), the young L7 dwarf PSO J318.5$-$22 (green; \citealp{faherty16}) and the old L7.5 dwarf Luhman 16A (blue; \citealp{faherty14}).  {\it Top:} Comparison of the two K I doublet FIRE orders, showing that \target\ and PSO J318.5$-$22 have comparably low surface gravities. {\it Bottom:} Full FIRE spectra, normalized to unity between 1.25-1.30$\micron$ and offset vertically by 0.5 units for clarity.  The $\lambda<1.1\micron$ region in the spectrum of \target\ has low SNR and the FIRE order stitching is unreliable. The high-order continuum change at $1.55\micron$ and the CO band head at $2.30\micron$ fall at FIRE order boundaries, so the strength of these features is also unreliable. The complete spectrum of PSO J318.5$-$22 (not shown) is presented in \cite{faherty16}.}
	\label{fig:youth}
\end{figure*}

To determine if \target\ is young, we focused our analysis on the gravity-sensitive K I absorption lines (1.1692 and 1.1778\,$\mu$m, 1.2437 and 1.2529\,$\mu$m). Our FIRE spectrum has 40 times the resolution of our previous SpeX prism spectrum \citep{2015AJ....150..182K}, allowing us to directly compare the strength of the spectral features to those of young and field-age late-L dwarfs. The top panels of Figure \ref{fig:youth} show sections of our FIRE spectrum (black) compared to FIRE spectra of the young L7 dwarf PSO~J318.5338$-$22.8603 (green; hereafter PSO~J318.5$-$22; \citealp{faherty16}) and the field-aged L7.5 dwarf Luhman 16A (blue; \citealp{faherty14}) centered on the K I lines. The K I lines in the \target\ spectrum are weaker than in Luhman 16A, indicating low surface gravity and hence a young age ($\lesssim$~200 Myr; \citealp{allers13}).  Conversely, the K I line strengths of \target\ and PSO J318.5$-$22 are indistinguishable from each other.  The two objects are among the reddest isolated L dwarfs known to date. The equivalent widths of the K I lines as defined by \cite{mclean03} for \target\ ($1.21 \pm 0.66$\,$\AA$ at 1.1168\,$\mu$m; $3.84 \pm 0.64$\,$\AA$ at 1.1177\,$\mu$m; $1.89 \pm 0.29$\,$\AA$ at 1.1243\,$\mu$m; and $3.12 \pm 0.31$\,$\AA$ at 1.1254\,$\mu$m) are systematically lower than those of field L7 dwarfs (e.g., see \citealp{mclean03}, Figure 15). The complete $R\sim6000$, 0.7$-$2.4 $\mu$m FIRE spectrum of \target\ is shown in the bottom panel of Figure \ref{fig:youth}, where it is seen to be significantly redder than the near-IR spectrum of the field-aged L7.5 dwarf Luhman 16A.

The combined evidence leads us to conclude that \target\ is indeed young, and comparable in age to the 23 Myr-old $\beta$ Pictoris association member PSO J318.5$-$22.  At the current signal-to-noise ratios and resolutions, we cannot determine if one object is younger than the other from the spectra alone. From this comparison, together with the very red color and the consistency of various spectral features and indices with other young, late-L dwarfs \citep{2015AJ....150..182K}, we conclude that \target\ has low surface gravity and is likely $\lesssim$~120 Myr old.

\section{KINEMATICS}\label{sec:rvmes}

We used single-exposure, single-order and telluric-corrected FIRE spectra to measure the radial velocity of \target. We used a forward modelling approach that relies on the zero-velocity BT-Settl atmosphere models \citep{2013MSAIS..24..128A,2003A&A...402..701B} to reproduce our observations. Our forward model employs 4 free parameters: a radial velocity shift, an instrumental line spread function (LSF) width where the LSF is modelled as a Gaussian function, and a two-parameters linear slope multiplicative correction to account for instrumental effects.

\begin{figure*}
	\centering
	\includegraphics[width=0.85\textwidth]{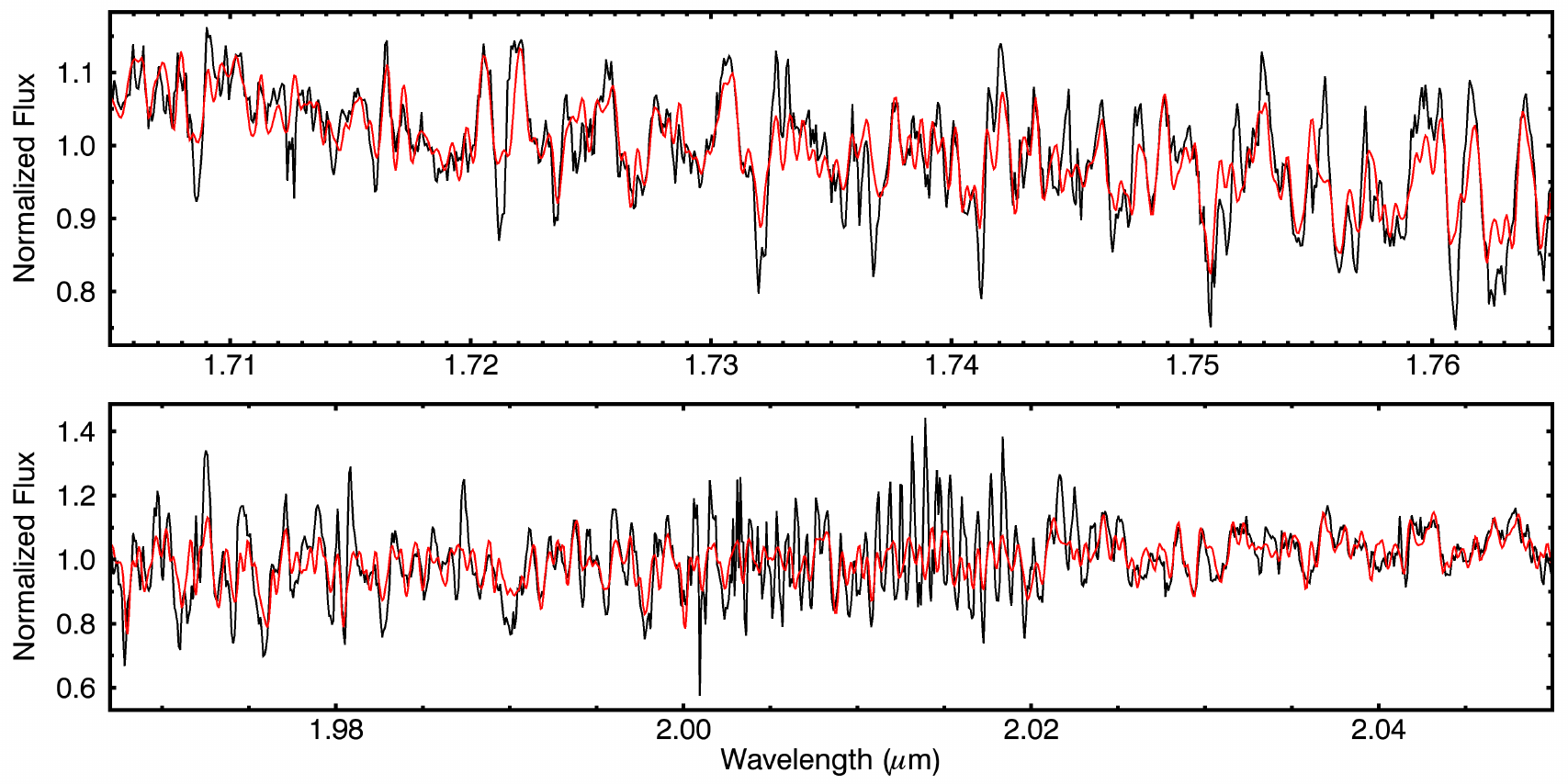}
	\caption{Forward model fitting (red line) to the observed spectra (black line) in echelle orders 15 and 13, in order to determine the radial velocity of \target.}
	\label{fig:modelfit}
\end{figure*}

We used the IDL implementation of AMOEBA to identify the set of parameters that minimize the reduced $\chi^2$ between the forward model and the observed spectrum. This is similar to the algorithm described by \cite{2016arXiv160104717A} except that telluric absorption was corrected a priori and we did not attempt to measure $v\sin i$ with the limited resolution of our data. This optimization was carried out on the 7 individual exposures, in a selected set of wavelength ranges that are located away from order boundaries and in regions where the S/N is high and sufficient radial velocity information is present in the science spectrum. We selected specific wavelength regions in orders 24 (1.092--1.120\,$\mu$m), 17 (1.510--1.554\,$\mu$m), 15 (1.680--1.790\,$\mu$m), and 13 (1.960--2.050\,$\mu$m), that satisfied these criteria. We used the BT-Settl model with \Teff\ = 1600\,K and $\log g = 4.0$, which yielded the smallest reduced $\chi^2$ in these orders with no velocity shift.  This temperature and surface gravity are in agreement with those found for PSO~J318.5$-$22 \citep{2013ApJ...777L..20L}.  We display two example model fits that we obtained in Figure~\ref{fig:modelfit}.

In each order, the useful wavelength range was split in 15 individual segments of width 0.02\,$\mu$m to account for sub-pixel systematic effects in the wavelength calibration (e.g., see \citealt{2015ApJ...808L..20G}). This yielded a total of 420 individual radial velocity measurements (7 exposures, 4 orders and 15 order segments). The single-exposure radial velocities were first combined at every wavelength range using an SNR-squared weighted mean. A weighted standard deviation was used to obtain measurement errors on each of these radial velocity measurements.

This method yielded a weighted-average radial velocity of $8.5 \pm 1.3$\,\kms. Similar radial velocity measurements with FIRE have been demonstrated to bear a 3\,\kms\ systematic uncertainty (e.g., see \citealt{2007ApJ...666.1205Z}), which we add in quadrature to our measurement to obtain $8.5 \pm 3.3$\,\kms. 

In order to calibrate our method, we applied it to FIRE data obtained for PSO~J318.5$-$22, for the young T5.5 SDSS~J111010.01+011613.1 \citep{2015ApJ...808L..20G}, and for three young M9.5--L2 brown dwarfs with Keck NIRSPEC radial velocity measurements (2MASS~J02340093--6442068, 2MASS~J02103857--3015313, 2MASS~J23225299--6151275; \citealp{faherty16}).  We found measurement discrepancies between 0.5\,\kms\ and 4.4\,\kms\ with a standard deviation of 1.5\,\kms, well within our measurement errors. These data will be detailed in a future publication. 

We solved for the proper motion from the 14-year baseline spanned by 2MASS, SDSS, WISE, WISE 3-Band Post-Cryo, and NEOWISE-R images, using relative astrometry as prescribed in \cite{kirkpatrick14}. The total proper motion differs slightly from the measurement in \cite{2015AJ....150..182K} which was obtained only from the difference between the 2MASS and AllWISE positions.

\section{TWA MEMBERSHIP}\label{sec:twa}

We used our radial velocity measurement along with the sky position and proper motion of \target\ (Table 1) to calculate a young moving group membership probability with the BANYAN~II tool\footnote{Publicly available at \url{http://www.astro.umontreal.ca/\textasciitilde gagne/banyanII.php}} \citep{2014ApJ...783..121G}. The BANYAN II tool assumes gaussian ellipsoid models adjusted to fit the XYZ and UVW distributions of known bona fide members of different moving groups, or the Besan\c{c}on galactic model \citep{robin12} for the field hypothesis. BayeÕs theorem is used to derive a probability density from the comparison of the measurements to the models, and the probability is obtained by marginalizing the probability density function over all possible distances.

\begin{deluxetable}{ll}
\tablecolumns{2}
\tablewidth{0pc}
\tablecaption{Characteristics of 2MASS~J11193254--1137466}
\startdata
Right Ascension & $11^h19^m32.543^s$ \\
Declination & $-$11\textdegree$37'46.70''$ \\
Radial Velocity & 8.5 $\pm$ 3.3 \kms\ \\
$\mu_\alpha\cos\delta$ & $-$145.1 $\pm$ 14.9 \masyr\ \\
$\mu_\delta$ & $-$72.4 $\pm$ 16.0 \masyr\ \\
$Y$ & 19.045 $\pm$ 0.093~mag \\
$J$ & 17.330 $\pm$ 0.029~mag \\
$H$ & 15.844 $\pm$ 0.017~mag \\
$K_s$ & 14.751 $\pm$ 0.012~mag \\
$W1$ & 13.548 $\pm$ 0.026~mag \\
$W2$ & 12.883 $\pm$ 0.027~mag \\
$\log L/L_{\odot}$ & $-$4.39 $\pm$ 0.14 \\
Age & 10 $\pm$ 3 Myr \\
Kinematic Distance & 28.9 $\pm$ 3.6 pc \\
Estimated Mass & 4.3$-$7.6 \MJup \\
\enddata
\tablecomments{ $YJHK_s$ magnitudes from the VHS.\footnote{\url{http://www.vista-vhs.org/}}}
\end{deluxetable}

\begin{figure}
	\centering
	\subfigure[$XYZ$ galactic position]{\includegraphics[width=0.55\textwidth]{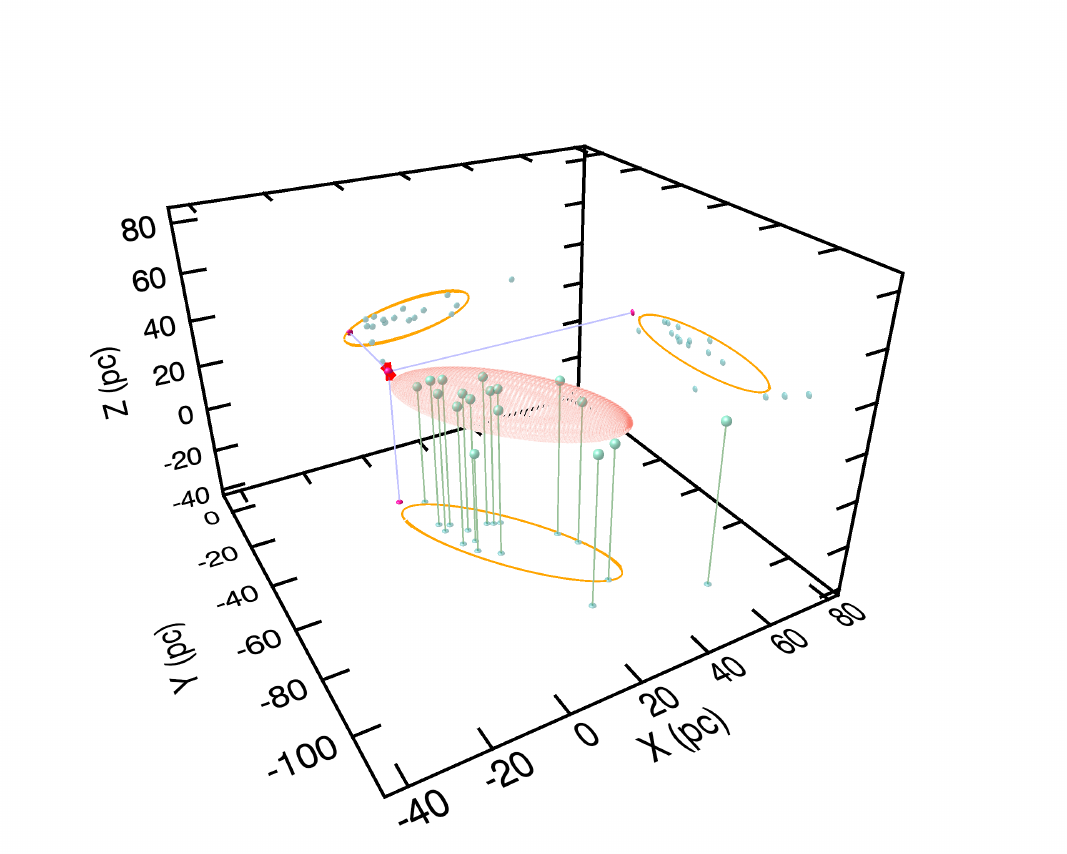}\label{fig:xyz}}
	\subfigure[$UVW$ space velocity]{\includegraphics[width=0.55\textwidth]{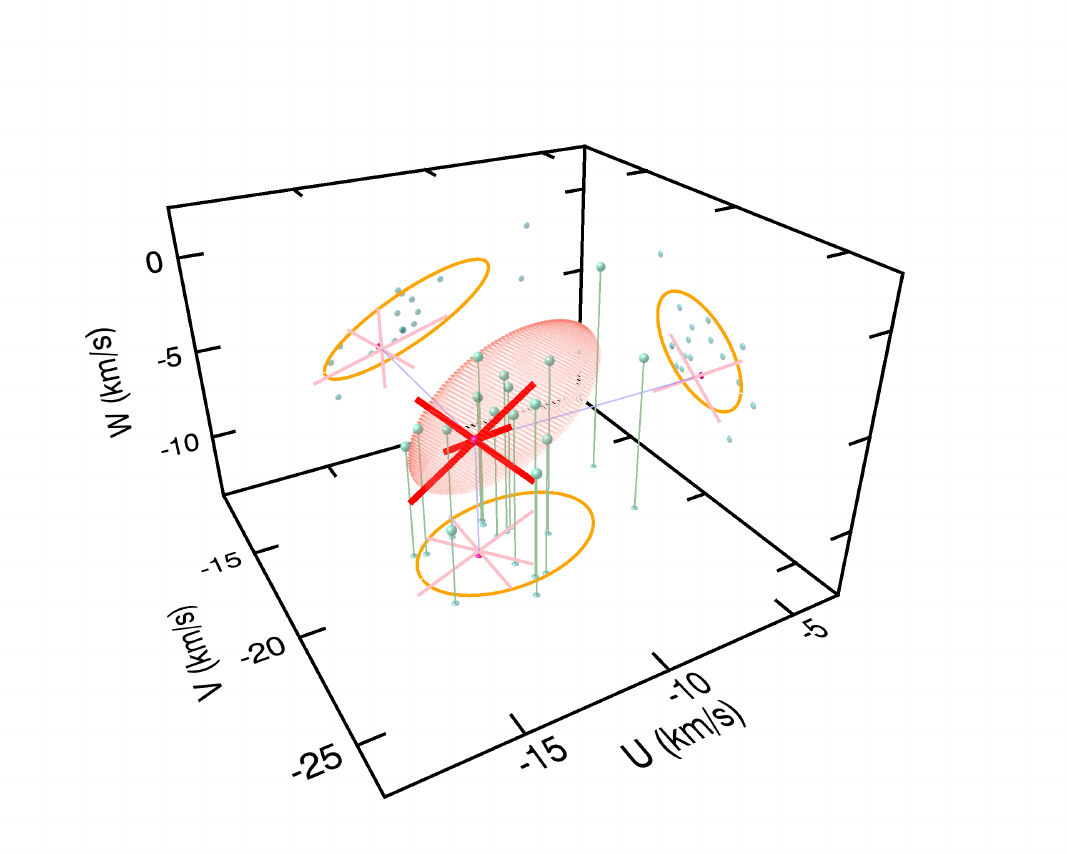}\label{fig:uvw}}
	\caption{Galactic position and space velocity of \target\ (red sphere and red error bars) compared to known TW~Hydrae bona fide members (green spheres) and the 1$\sigma$ Gaussian ellipsoid spatial and kinematic models used in BANYAN~II (orange ellipses). The data are projected to the three 2D planes for an easier visualization.}
	\label{fig:xyzuvw}
\end{figure}

We obtained an 88\% probability that \target\ is a TWA member, with a calibrated field contamination probability of 0.003\% equivalent to a statistical significance of 4.15$\sigma$.  The coordinate that puts the most stringent constraint on the membership of \target\ is Z, which only matches the distribution of TWA members. The statistical kinematic distance is $28.9 \pm 3.6$\,pc, consistent with our earlier photometric estimate of $27 \pm 7.2$\,pc \citep{2015AJ....150..182K} which was calculated from the apparent $K_s$ magnitude and the expected absolute $K_s$ magnitude of a young L7 dwarf \citep{filippazzo15}. The distance is also consistent with the absolute magnitude relations for other young, late-type L dwarfs (e.g. WISEP~J004701.06+680352.1 and PSO~J318.5$-$22; \citealp{faherty16}), which take into account all young L dwarfs. When we include the photometric distance estimate as an input to the BANYAN~II algorithm, the TWA membership probability increases to 92\% and the field contamination probability significantly decreases to 0.0005\%, meaning that \target\ is consistent with being a TWA member at a statistical significance of 4.55$\sigma$.

In Figure~\ref{fig:xyzuvw} we display the galactic position $XYZ$ and space velocity $UVW$ of \target\ at its statistical distance, compared to known bona fide TWA members, and to the 1$\sigma$ contours of the 3D Gaussian ellipsoid models used in BANYAN~II. Although the current model of TWA extends its 1$\sigma$ contour to only $\sim$33 pc, this $\sim$25 pc object appears to have been drawn from the same gaussian random distribution in $XYZUVW$ space. The $XYZ$ position of \target\ reveals it to be the nearest TWA member and extends the observed distance distribution of currently known objects. While other closer candidate TWA members have been reported in the literature (TWA 22; \citealp{song03}), their space positions and velocities fall well outside of the TWA ellipsoid \citep{mamajek05,teixeira09, weinberger13}.  Our own calculation with BANYAN II yields an association probability of 10$^{-7}$ for TWA 22.

At the age of TWA ($10 \pm 3$\,Myr; \citealt{2015MNRAS.454..593B}), we estimate that the mass of \target\ is 4.3--7.6\,\MJup, where we have compared the bolometric luminosity derived from the $K_s$ magnitude at its TW Hydrae kinematic distance and age to model predictions from \cite{saumon08} and directly compared the absolute $K_s$ magnitude to the predictions from \cite{baraffe15}.

The mass makes \target\ comparable to the early-L companion object, 2M1207b \citep{chauvin04,chauvin05}, which is also a member of TW Hydrae \citep{gizis02}. The two objects have similar $H-K_s$ colors ($1.16\pm0.24$~mag for 2M1207b; \citealp{chauvin04}) to within the uncertainties, although 2M1207b is overall redder in the 1--4~$\micron$ region \citep{chauvin04,song06}.  The 0.4~dex lower luminosity of 2M1207b \citep{filippazzo15} reveals that it is likely lower in mass, regardless of model assumptions.  Comparing to the cooling tracks of \cite{saumon08} and \cite{baraffe15}, 2M1207b has an estimated mass of 3.3--5.4 \MJup.  Our mass estimate for \target\ is slightly higher, 4.3--7.6 \MJup, which still leaves it as the least massive isolated member of TWA.  Both objects are at or above the theoretical opacity-limited turbulent fragmentation limit of 3--5 \MJup\ \citep{boyd05}, so both could have formed through star-like gravo-turbulent collapse.  However, both have masses comparable to or smaller than the 5--7 \MJup\ masses of the HR 8799 planets \citep{marois08,marois10}. Hence both \target\ and 2M1207b could have in principle formed, and in the case of \target\ subsequently been ejected, via a planet-like mechanism.

We sought bright common proper motion companions to \target\ by calculating the 2MASS--AllWISE proper motions of all 2MASS entries within 60$'$, corresponding to a maximum physical separation of 22,500\,AU at 25\,pc. None of the 368 2MASS sources within this radius had a proper motion within 76\,\masyr\ of that of \target, indicating no potential stellar host to this young planetary-mass object.

\section{CONCLUSION}\label{sec:conclusion}

We have spectroscopically confirmed the low surface gravity, and hence youth, of the L7 dwarf \target. Its radial velocity, proper motion, and galactic position are consistent with that of the $10 \pm 3$ Myr-old TWA.  From the object's near-infrared absolute magnitudes we determine a mass of 4.3--7.6\,\MJup.  It is the nearest member of TWA, and the second-brightest isolated $<$10\,\MJup\ object discovered to date.  Hence, \target\ is an excellent benchmark for young, directly imaged extrasolar planets.

\acknowledgements

We thank Adam Burgasser and Katelyn Allers for helpful comments with respect to the radial velocity measurement method. We thank Robert Simcoe for help with the data reduction. This work was supported by an NSERC Discovery grant to S.M.\ at the University of Western Ontario. This work was performed in part under contract with the California Institute of Technology (Caltech)/Jet Propulsion Laboratory (JPL) funded by NASA through the Sagan Fellowship Program executed by the NASA Exoplanet Science Institute. This paper includes data gathered with the 6.5 meter Magellan Telescopes located at Las Campanas Observatory, Chile.

\facility{Gemini-South (Flamingos-2); Magellan:Baade (FIRE)}
\software{IDL, FIREHOSE, IRAF}

\bibliographystyle{apj}

\end{document}